\documentclass[preprint,12pt]{elsarticle}

\usepackage{epsfig}

\usepackage{subcaption}
\usepackage{siunitx}
\usepackage[colorlinks]{hyperref}
\usepackage[labelfont=bf]{caption}

\usepackage{amssymb}
\usepackage{amsthm}

\usepackage{amsmath}

\usepackage{xcolor}
\usepackage{float}

\DeclareUnicodeCharacter{03B3}{$\gamma$}
\DeclareUnicodeCharacter{2032}{\ensuremath{'}}
\DeclareUnicodeCharacter{2215}{\ensuremath{/}}

\usepackage{color}

\journal{Journal of Computational Physics}

\begin{document}

\begin{frontmatter}


\title{Dislocation dynamics in Ni-based superalloys: Parameterising dislocation trajectories from atomistic simulations}

\author[1]{Geraldine Anis\corref{cor1}}
\ead{Geradine.Anis@warwick.ac.uk}
\author[2]{Thomas Hudson}
\author[3]{Peter Brommer}
\ead{p.brommer@warwick.ac.uk}

\cortext[cor1]{Corresponding author}

\affiliation[1]{organization={EPSRC Centre for Doctoral Training for Modelling of Heterogeneous Systems, University of Warwick},
            city={Coventry},
            postcode={CV4 7AL}, 
            country={UK}}

\affiliation[2]{organization={Warwick Mathematics Institute, University of Warwick},
            city={Coventry},
            postcode={CV4 7AL}, 
            country={UK}}

\affiliation[3]{organization={Warwick Centre for Predictive Modelling, School of Engineering, University of Warwick},
            city={Coventry},
            postcode={CV4 7AL}, 
            country={UK}}

\begin{abstract}

Nanoscale precipitates in the microstructure of nickel-based superalloys hinder dislocation motion, which results in an extraordinary strengthening effect at elevated temperatures. 
We used molecular dynamics (MD) with classical effective potential to observe the movement of an $\frac{a}{2}\langle110\rangle\{111\}$ edge dislocation under shear in pure Ni, which  represents the Ni solid solution matrix, and extracted the locations of the dislocations. 
We show how a Differential Evolution Monte Carlo (DE-MC) analysis is an effective way to find the parameters of an equation of motion for the dislocation lines with quantified uncertainties. 
The parameters of interest were the effective mass, drag coefficient, and force experienced by the dislocation. 
The marginal parameter and joint posterior distributions were estimated from the accepted samples produced by the DE-MC algorithm. 
The equation of motion and parameter distributions were used to predict the dislocation positions and velocities at the simulation timesteps, and the mean fit was found to match the MD trajectories with a root mean square error (RMSE) of \SI{0.2}{\nano\metre}. 
We also discuss how the selected model can be extended to account for the presence of multiple dislocations as well as dislocation-precipitate interactions. 
This work serves as the first step towards building a predictive surrogate model that describes the deformation behaviour of Ni-based superalloys.
\end{abstract}

\begin{keyword}
Dislocations \sep Molecular dynamics \sep Ni-based superalloys \sep Surrogate modelling \sep Uncertainty quantification

\end{keyword}

\end{frontmatter}

\section{Introduction}
\label{sec:Introduction}
Ni-based superalloys are important materials that are widely used in high-temperature applications. 
Precipitation strengthening by $\gamma'$ (Ni$_3$Al) precipitates, which hinder dislocation motion, plays a major role in providing the extraordinary strength exhibited by these alloys. 
The ordered L1$_2$ $\gamma'$ phase is precipitated out of the face-centered cubic (FCC) $\gamma$ matrix, which is a Ni solid solution phase.

Understanding the dynamics of dislocations in these alloys is central to studying their deformation behaviour. 
Here, we present a methodology to parameterise the motion of dislocations in the $\gamma$ matrix based on the work of Bitzek \emph{et al.}\  \cite{bitzek_atomistic_2004, bitzek_atomistic_2004-1, bitzek_dynamic_2005}. 
Pure FCC Ni is used as a model system to represent the $\gamma$ phase in these alloys. The effect of the $\gamma'$ precipitates on the dislocation motion is considered in a future work. 
In FCC metals, the primary dislocations of interest are the $\frac{a}{2}\langle110\rangle\{111\}$ edge dislocations, which are perfect dislocations. 
It is energetically favorable for these dislocations to dissociate into two partial dislocations with a stacking fault (SF) in between:

\begin{equation}
\frac{1}{2}[\bar110] \rightarrow \frac{1}{6}[\bar211] + SF + \frac{1}{6}[\bar12\bar1]. \label{eqn:disloc_dissociation}
\end{equation}

Previously, Bitzek \emph{et al.}\ \cite{bitzek_atomistic_2004, bitzek_atomistic_2004-1, bitzek_dynamic_2005} postulated an equation of motion (\ref{eqn:dim_model}) of the form 
\begin{equation}
m\frac{d^2x}{dt^2} + B\frac{dx}{dt} = F,\label{eqn:dim_model}
\end{equation}
and identified the parameters through Least Squares (LS) fitting to dislocation trajectories generated using molecular dynamics (MD). 
In equation (\ref{eqn:dim_model}), $m$ is the effective mass of the dislocation and determines the dislocation's effective inertia; $B$ is the drag coefficient and describes the lattice resistance to dislocation motion; and $F$ is the force experienced by the dislocation. 
$F$ is assumed to decompose as $F = F_a + F_i$ , where $F_a$ is the Peach-Koehler force \cite{peach_forces_1950} and $F_i$ is the force that results from dislocation interactions with defects and lattice resistance. 
It should be noted that at dislocation velocities comparable with the sound speed in the material, the effective dislocation mass $m$ has a velocity-dependent component such that $m_{\text{eff}} = m_0 + f(v)$, where $m_0$ is the rest mass of the dislocation and $f(v)$ is a function of the dislocation velocity $v$ \cite{sakamoto_high-velocity_1991}. 
In the following, the velocity dependence of the mass is ignored and our analysis is limited to slow-moving dislocations under low applied shear stresses.

Here, rather than using LS, we use Differential Evolution Monte Carlo (DE-MC) \cite{cajo_j_f_ter_braak_markov_2006, cajo_j_f_ter_braak_differential_2008} as a sampling approach to fit the model parameters in (\ref{eqn:dim_model}) to edge dislocation trajectories in pure Ni. 
DE-MC is an adaptive Monte Carlo technique with the advantage of fast convergence in cases where parameters are highly correlated, which is the case for the current model. Using a statistical approach to parameter fitting like DE-MC enables us to quantify the uncertainty in the resulting model based on the data available, as we discuss further below.
In conventional DE-MC, several Markov chains are run in parallel and the current state of a given chain, $\boldsymbol{\theta}_i$, which corresponds to a parameter vector, is updated by generating a proposal, $\boldsymbol{\theta}_p$, which depends on the states of two randomly selected chains from the pool of remaining chains, $R_1$ and $R_2$: 
\begin{equation}
\boldsymbol{\theta}_p = \boldsymbol{\theta}_i + \gamma (\boldsymbol{\theta}_{R_1} - \boldsymbol{\theta}_{R_2}) + \mathbf{e} \label{eqn:de_mc_1},    
\end{equation}
  where $\gamma$ is a scaling factor and $\mathbf{e}$ is a random vector drawn from a narrow, symmetric distribution. 
  A proposal, $\boldsymbol{\theta}_p$, is accepted with a probability min(1, $r$), where $r = \pi(\boldsymbol{\theta}_p)/\pi(\boldsymbol{\theta}_i)$ and $\pi(.)$ is the probability density function (pdf) of the $d$-dimensional target distribution to be sampled.
  If past states of chains are considered for the update, this is denoted as DE-MC$_Z$. DE-MC$_{ZS}$ \cite{cajo_j_f_ter_braak_differential_2008} (used here), uses both past states as well as a snooker update, where a given chain is updated according to the following:
\begin{equation}
    \boldsymbol{\theta}_p = \boldsymbol{\theta}_i + \gamma (\mathbf{z}_{P1} - \mathbf{z}_{P2}),
    \label{eq:DEMZ_ZS}
\end{equation}
where $\mathbf{z}_{P1}$ and $\mathbf{z}_{P2}$ are the orthogonal projections of $\boldsymbol{\theta}_{R1}$ and $\boldsymbol{\theta}_{R2}$ from (\ref{eqn:de_mc_1}) on to the line $\boldsymbol{\theta}_i - \mathbf{z}$, where $\mathbf{z}$ is the state of another chain. The acceptance ratio in this case then becomes:
\begin{equation}
    r = \frac{\pi (\boldsymbol{\theta}_p) ||\boldsymbol{\theta}_p - \mathbf{z}||^{d-1}} {\pi (\boldsymbol{\theta}_i)||\boldsymbol{\theta}_i - \mathbf{z}||^{d-1}}.
\end{equation}

Both the DE-MC$_Z$ and DE-MC$_{ZS}$ algorithms reduce the number of chains needed for the DE-MC sampling to converge.
For more information on the development of the DE-MC algorithm and the more recent snooker update, the reader is referred to Refs.\ \cite{cajo_j_f_ter_braak_markov_2006} and \cite{cajo_j_f_ter_braak_differential_2008}.

Parameterising the motion of dislocations in this way is useful, as the resulting parameters can be used as inputs to larger lengthscale approaches (e.g. discrete dislocation dynamics (DDD) \cite{bitzek_atomistic_2004}), which are needed to capture the complex nature of the deformation behavior of metals. 
Moreover, there are several advantages to using sampling to fit the model in (\ref{eqn:dim_model}) to dislocation position data. 
Using sampling, we not only get parameter values but also their probability distributions. 
This means that we are able to select the most probable parameter values as well as obtaining parameter distributions that allow us to quantify the uncertainty in the model predictions, namely the dislocation positions and velocities.

In Sec.\ \ref{sec:methods} we describe the methods underlying the research, mainly MD and DE-MC. 
This is followed by a presentation of the results and their discussion in Sec.\ \ref{sec:results}. 
The main conclusions can be found in Sec.\ \ref{sec:conclusions}.

\section{Methods}\label{sec:methods}
\subsection{Molecular dynamics simulations of edge dislocations}
\begin{figure}[h]
\centering
\includegraphics[width=4in]{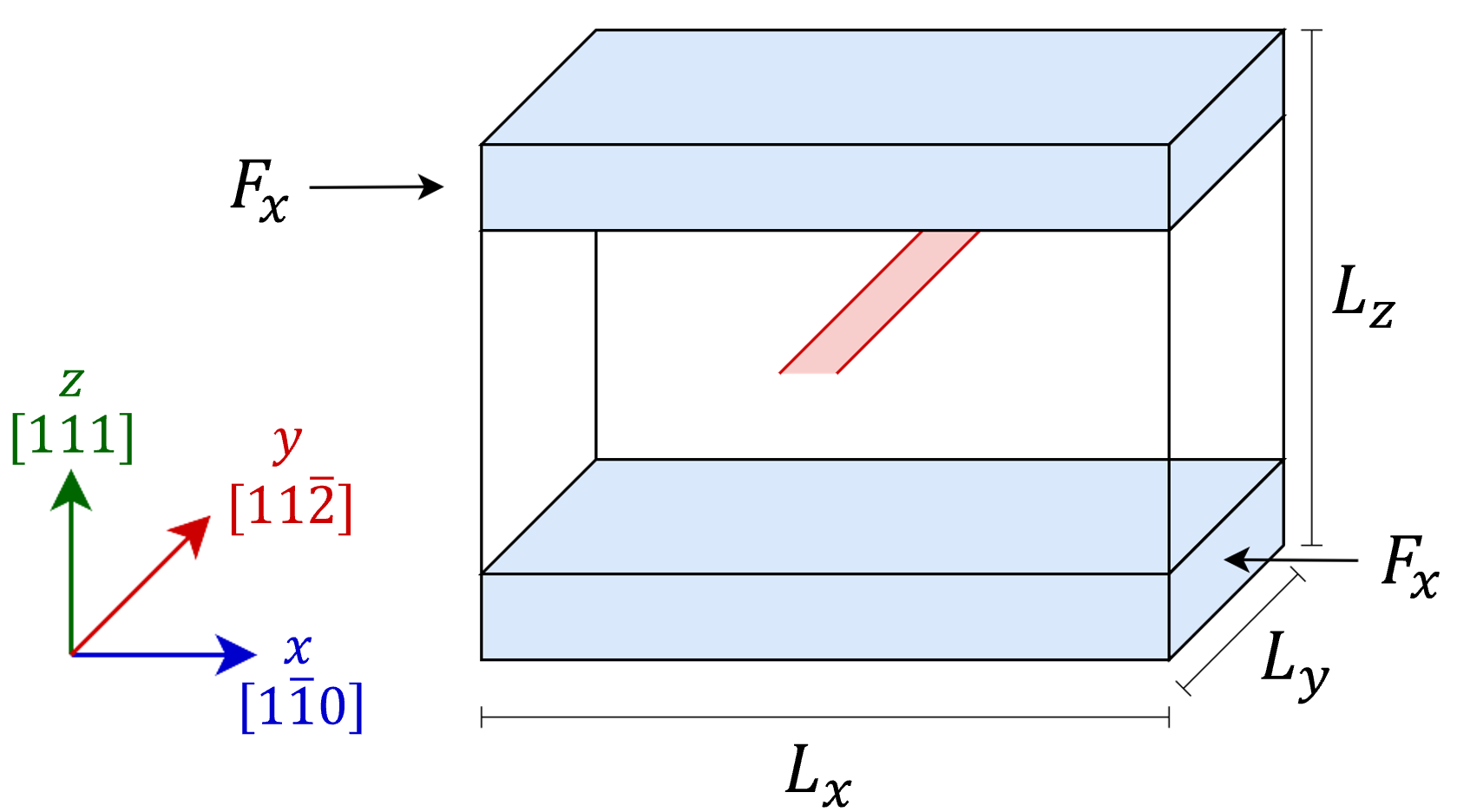}
\caption{Schematic of the simulation box used. 
The dissociated edge dislocations is shown as the two $\frac{a}{6}<112>$ partial dislocations in red with the stacking fault in between. 
Periodic boundaries are applied in the $x$ and $y$ directions. 
The cell is non-periodic in $z$. 
To shear the cell, a force $F_x$ is applied to the atoms in the top and bottom layers of cell shaded in blue.}
\label{fig:sim_box}
\end{figure}

Using the LAMMPS package \cite{thompson_lammps_2022}, Molecular Dynamics (MD) simulations of $\frac{a}{2}\langle 110\rangle\{111\}$ edge dislocations in pure Ni were carried out using the Mishin 2004 \cite{mishin_atomistic_2004} embedded atom method (EAM) \cite{daw_embedded-atom_1984} potential. 
The lattice constant, $a_0$, used for Ni was \SI{0.352}{\nano\metre} and the MD timestep was \SI{0.001}{\pico\second}. 
To create the edge dislocation, a single [110] plane was removed from a pure Ni block, followed by relaxing the atom positions to get the correct dislocation core structure.
This leads to the dissociation of the dislocation into two partial dislocations as is expected for FCC materials. 
The dislocation was inserted in a box with dimensions $L_x = \SI{38}{\nano\metre}$, $L_y = \SI{17}{\nano\metre}$, and $L_z = \SI{30}{\nano\metre}$, which was selected after testing a range of box sizes to avoid finite size effects.
Periodic boundary conditions (PBCs) were applied in the $[1\bar10]$ direction (the glide direction of the dislocation) as well as the $[11\bar2]$ direction (along the dislocation line) such that an infinitely long dislocation is obtained. 
A schematic of the simulation box used is shown in Fig.\ \ref{fig:sim_box}.

To simulate the dynamics of the edge dislocation, the simulation box was deformed under a shear stress, $\sigma = \SI{10}{\mega\pascal}$, by applying a constant force $F_x =\pm \sigma A/N$ to the atoms in the six top and bottom atomic layers of the box, where $A = L_x \times L_y$ is the cross-sectional area of the simulation box in the $xy$-plane and $N$ is the number of atoms in each layer. 
To reduce the effects of the initial stress wave, the applied force was first linearly ramped from 0 to the target value $F_x$ over a period of \SI{20}{\pico\second} \cite{cereceda_techniques_2012}. 
Following this initial force ramping, the constant force $F_x$ was applied to the simulation box for \SI{100}{\pico\second} and only results from this portion of the simulation were used in the analysis. 
The simulations were carried out in the canonical ensemble at \SI{300}{\kelvin} using a Nos\'e-Hoover thermostat \cite{shinoda_rapid_2004}, where the system was allowed to equilibrate prior to applying the shear force.

The OVITO Dislocation Analysis Tool (DXA) \cite{stukowski_automated_2012} was used to extract the positions of the partial dislocations from the \SI{100}{\pico\second} MD trajectories. 
The positions of the two partial dislocations were averaged to study the dynamics of the perfect dislocation. 
To get statistics on the dislocation positions, 8 identical simulations were carried out according to the setup described above with a different random seed used for generating the atom starting velocities each time. 
The dislocation positions, extracted using DXA, were then averaged across the different realisations to get the mean dislocation position at each timestep and the corresponding standard deviation was also calculated. The raw dislocation trajectories used in the analysis are shown in green in Fig. \ref{fig:dim_fit}, and the position-time data is provided in the accompanying dataset \cite{geraldine_anis_dataset_nodate}.

\subsection{Model fitting using DE-MC}

After obtaining the mean dislocation positions as a function of time, the parameters $m$, $B$ and  $F$ were determined by fitting equation (\ref{eqn:dim_model}) to the position data using DE-MC. 
Prior to fitting, some manipulation of equation (\ref{eqn:dim_model}) was necessary; dividing through by $m$, we obtain
\begin{equation}
\frac{d^2x}{dt^2} + \frac{B}{m} \frac{dx}{dt} = \frac{F}{m}. \label{eqn:dim_model_1}    
\end{equation}
The choice to scale the coefficients in this way was made to avoid the infinite family of $m$, $B$ and $F$ that generate identical solutions to equation (\ref{eqn:dim_model}).
In order to recover the full set of parameters, at low velocities relative to the speed of sound, the drag coefficient, $B$, can be calculated from
\begin{equation}
B = \frac{\tau b}{v_t} \label{eqn:B}   
\end{equation}
for a given applied shear stress $\tau$, Burgers vector $b$, and the terminal (equilibrium) dislocation velocity $v_t$. 
The value of $B$ calculated in this way can then used with the fitted $\frac{B}{m}$ and $\frac{F}{m}$ to calculate the corresponding values of $m$ and $F$.

Equation (\ref{eqn:dim_model_1}) was then put in a non-dimensional form:
\begin{equation}
\frac{d^2\tilde{x}}{d\tilde{t}^2} + c_0 \frac{d\tilde{x}}{d\tilde{t}} = c_1 \label{eqn:non_dim_model}
\end{equation}
 using the expressions for the non-dimensional positions $\tilde x$, and time $\tilde t$,
 \begin{equation}
\tilde{x} = \frac{x-x_0}{x_s},\hspace{1em}
\tilde{t} = \frac{t}{t_s},\hspace{1em}
\label{eqn:non_dim} \\ 
\end{equation}
where $x_0$ is the initial position of the dislocation, $x_s$ is the range of the mean dislocation position over \SI{100}{\pico\second}, and $t_s$ is the total simulation time. The resulting non-dimensional parameters $c_0$ and $c_1$ are related to the ratios $\frac{B}{m}$ and $\frac{F}{m}$, respectively, by the following expressions:
\begin{equation}
c_0 = \frac{B t_s}{m},\hspace{1em}
c_1 = \frac{F t_s^2}{m x_s}.\hspace{1em}
\label{eqn:non_dim_params}    
\end{equation}
Finally, to ensure the non-dimensional parameters are positive, $c_0$ and $c_1$ are represented as exponentials of scalar parameters $\tilde{c}_0$ and $\tilde{c}_1$, giving
\begin{equation}
\frac{d^2\tilde{x}}{d\tilde{t}^2} + e^{\tilde c_0} \frac{d\tilde{x}}{d\tilde{t}} = e^{\tilde c_1}, \label{eqn:fit}
\end{equation}
with $\tilde c_i=\ln c_i$.

DE-MC with twelve chains was then used to fit $\tilde c_0$ and $\tilde c_1$ in equation (\ref{eqn:fit}) to the non-dimensionalised position vs.\ time data obtained from MD simulations of edge dislocations in pure Ni. 
The $\text{MC}^3$ code \cite{cubillos_correlated-noise_2016} was used for the implementation of the DE-MC algorithm (with the snooker updater, referred to as DE-MC$_{ZS}$) within a Bayesian framework. 
Given the MD dislocation position data denoted by $\mathbf{y} = (y_1, \ldots, y_n)$, the target distribution is the joint probability density function (PDF), $p(\boldsymbol{\theta} | \mathbf{y}, \mathbf{t})$, where $y_i$ is the average position of the perfect dislocation at time, $t_i$, such that $\mathbf{t} = (t_1, ..., t_n)$, and $\boldsymbol{\theta}$ is the vector of model parameters, $\boldsymbol{\theta} = (\theta_0 \,, \theta_1)$. 
Assuming a Gaussian noise $\epsilon_i \sim \mathcal{N}(0,\sigma_i^2)$, the data can be modeled as the sum of the model predictions, $f_i(t_i;\theta)$, and the noise, $\epsilon_i$:
\begin{equation}
y_i = f_i(t_i; \boldsymbol{\theta}) + \epsilon_i \label{eqn:data}.    
\end{equation}
The posterior distribution of the parameters, $\theta$,  $p(\theta | \mathbf{y}, \mathbf{t})$, is then proportional to the product of a Gaussian likelihood and the prior:

\begin{eqnarray}
p(\theta | \mathbf{y}, \mathbf{t}) &=& p(\mathbf{y}|\mathbf{\boldsymbol{\theta}}) \times \text{Prior} \nonumber \\
&=&\mathcal{N}\left(\mathbf{y}|f(\mathbf{t};\boldsymbol{\theta}),\boldsymbol{\sigma}^2\mathbf{I}\right) \, \times \, \text{Prior} \nonumber \\
&=& \prod\limits_i \mathcal{N}\left(y_i | f(t_i;\boldsymbol{\theta}), \sigma_i\right) \, \times \, \text{Prior} \nonumber \\
&=& \prod\limits_i \left[\frac{1}{\sigma_i\sqrt{2\pi}} \, \text{exp} \left(\frac{f(t_i;\boldsymbol{\theta}) - y_i}{2\sigma_i}\right)^2 \right] \times \text{Prior} \label{eqn:posterior}    
\end{eqnarray}

In the $\text{MC}^3$ code \cite{cubillos_correlated-noise_2016}, the logarithm of the posterior distribution, $\log p(\boldsymbol{\theta} | \mathbf{y}, \mathbf{t})$, is used. Ignoring the constant terms, $\log p(\boldsymbol{\theta} | \mathbf{y}, \mathbf{t})$ is given by:

\begin{equation}
\log p(\boldsymbol{\theta} | \mathbf{y}, \mathbf{t}) = \sum\limits_i \left(\frac{f(t_i;\boldsymbol{\theta}) - y_i}{2\sigma_i}\right)^2  + \log \text{Prior}
\end{equation}

For the results presented here, a prior distribution $\mathcal{N}(0.0,2.0^2)$ was used for each of the parameters. 
The noise, $\epsilon_i \sim \mathcal{N}(0,\sigma_i^2)$, is assumed to be uncorrelated, where $\sigma_i$ is the standard deviation calculated from the dislocation positions at each timestep across the different simulation replicas. 
This corresponds to an assumption of heteroskedastic noise.
The DE-MC was run for $10^5$ generations and the accepted samples were used to estimate the posterior distribution by calculating the mean of each parameter and the covariance between them. 
A burn-in of $10^4$ samples were discarded and were not included in the analysis. 
To evaluate the convergence of the sampling, trace plots as well as convergence plots of the parameter means and standard deviations with the number of DE-MC generations were used.  

To quantify the uncertainty in the dislocation positions predicted by the model, samples were drawn from the estimated posterior distribution and used to evaluate the model at each timestep. 
Accordingly, a distribution was obtained for the dislocation position at every time $t_i$. 
A mean fit to the position data was obtained by evaluating the model using the mean parameter values, and the uncertainty in the position predictions was calculated as $\pm 2$ standard deviations of the distributions obtained at each time $t_i$ from the posterior distribution samples.
The root mean squared error (RMSE) of the fit was calculated by comparing the mean predicted positions to the mean positions obtained using MD. 

The marginal posterior distributions of the fitted parameters are also reported. 
When doing the sampling using DE-MC, if $\mathbf{\theta} = (\theta_0 \,, \theta_1)$ is an accepted sample, then $c_i = e^{\theta_i}$ and $\ln c_i = \theta_i$. 
Accordingly, in our analysis, the marginal posterior distributions presented are for $\ln c_i$, since these are normally distributed and are convenient to work with.
However, when showing the model fit to the position data, the dimensional forms $m$, $B$, and $F$ are recovered and the mean values are reported for these quantities.

\section{Results and discussion}\label{sec:results}
\subsection{Marginal parameter and joint posterior distributions}

\begin{figure}[h]
\centering
\includegraphics[width=.95\textwidth]{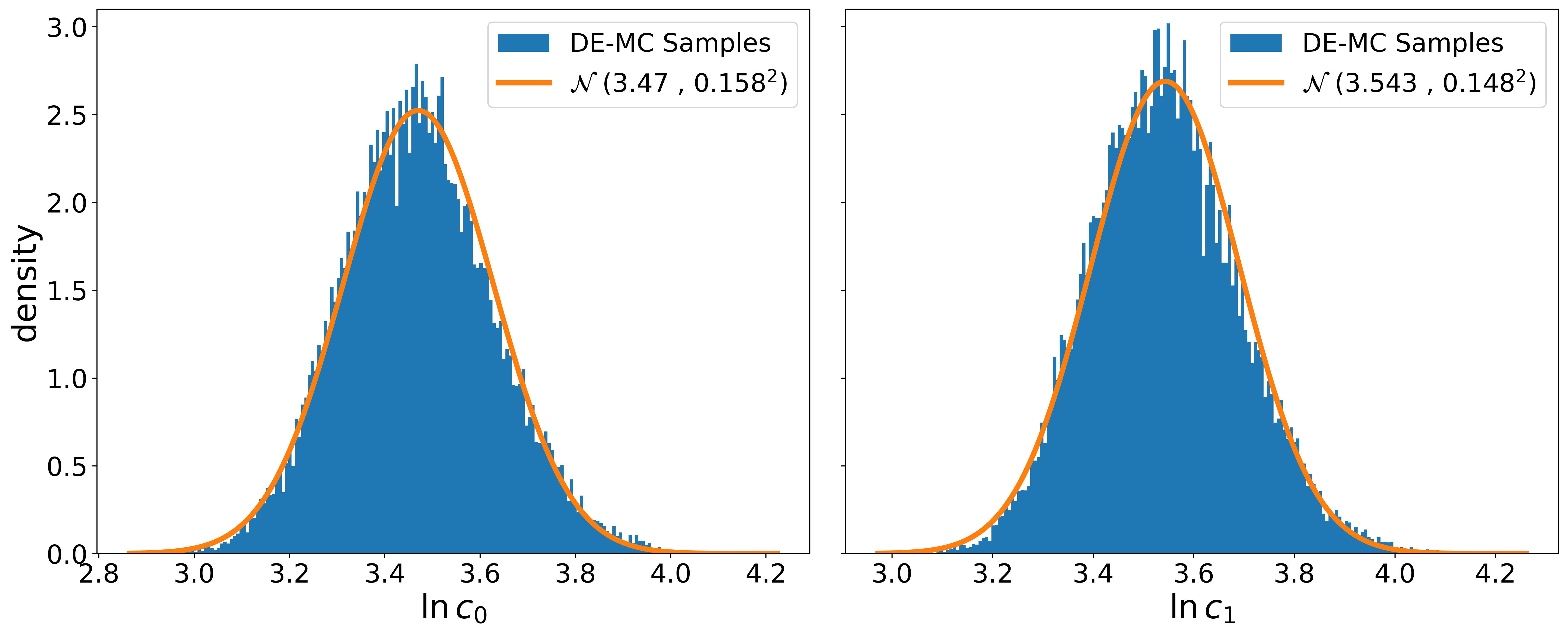}
\caption{Marginal posterior distributions of $\ln c_0$ and $\ln c_1$ estimated from the accepted samples from the DE-MC sampling.}
\label{fig:marginal_posterior}
\end{figure}

Figure \ref{fig:marginal_posterior} shows the histograms of the accepted samples for $\ln c_0$ and $\ln c_1$, normalised to give the probability density. 
Using the accepted samples, the mean and standard deviation of each parameter were calculated, and the parameters were found to follow $\ln c_0 \sim \mathcal{N}(3.470, 0.158^2)$ and $\ln c_1 \sim \mathcal{N}(3.543, 0.148^2)$. These are the marginal parameter posterior distributions estimated from the accepted samples, which are also shown in Fig.\ \ref{fig:marginal_posterior} and fit the histograms well.
The joint posterior distribution is given by the multivariate normal distribution, $\mathcal{N}(\boldsymbol{\mu}, \boldsymbol{\Sigma})$, with mean vector, $\boldsymbol{\mu}$, and covariance matrix, $\boldsymbol{\Sigma}$, which was also calculated from the accepted samples. 
These are given as: 
\begin{equation}
    \mathcal{N}(\boldsymbol{\mu} ,\, \boldsymbol{\Sigma}) ,\,
    \hspace{1em}
    \boldsymbol{\mu} =  
    \begin{pmatrix}
        3.470 \\
        3.543
    \end{pmatrix} ,\,
    \hspace{1em}
    \boldsymbol{\Sigma} =
    \begin{pmatrix}
        0.025 & 0.023 \\
        0.023 & 0.022
    \end{pmatrix}
\label{eqn:estimated_posterior}    
\end{equation}

\begin{figure}[ht]
\centering
\includegraphics[width=.80\textwidth]{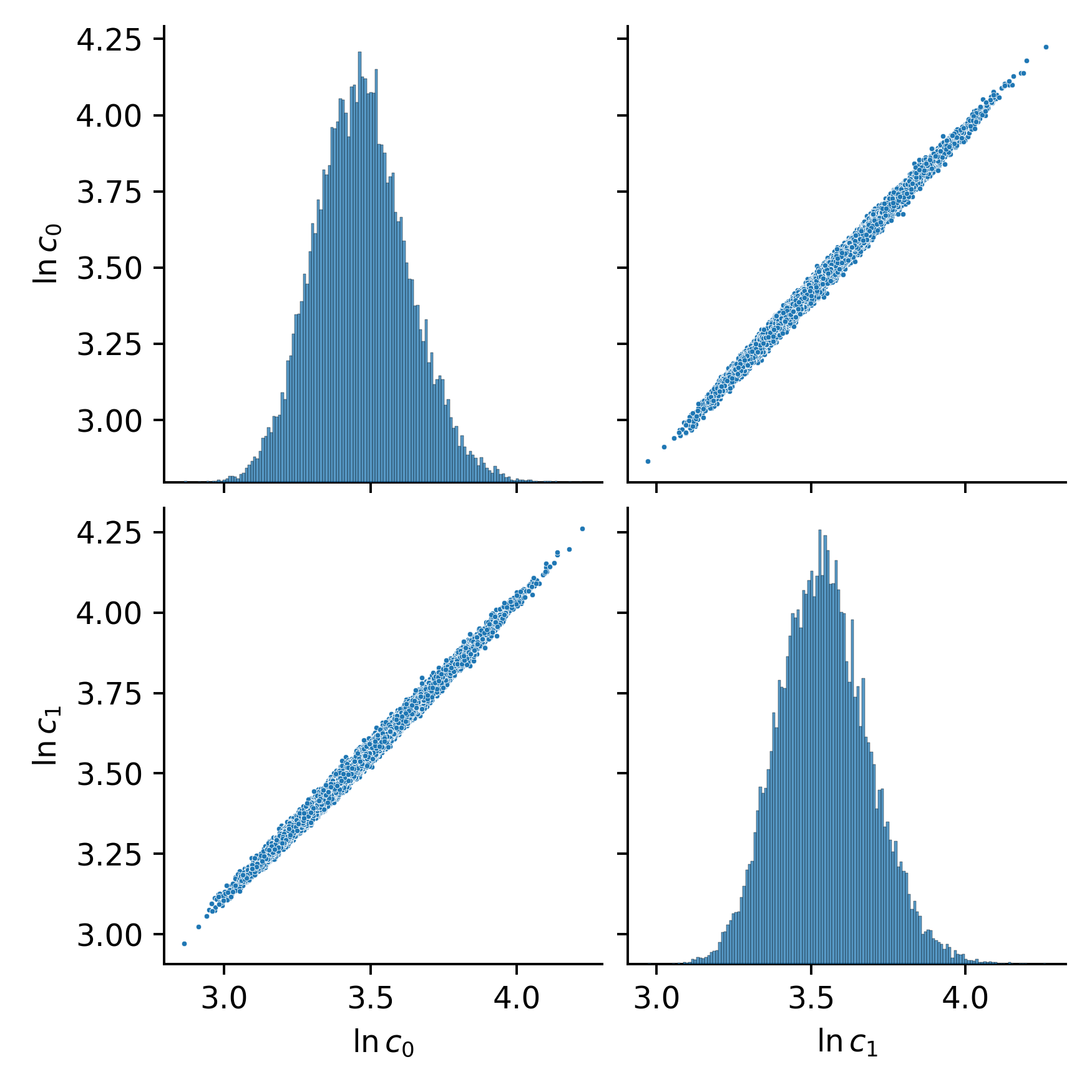}
\caption{Pair plot showing the correlation between $\ln c_0$ and $\ln c_1$.}
\label{fig:pairplot}
\end{figure}

\begin{figure}[ht]
\centering
\includegraphics[width=.95\textwidth]{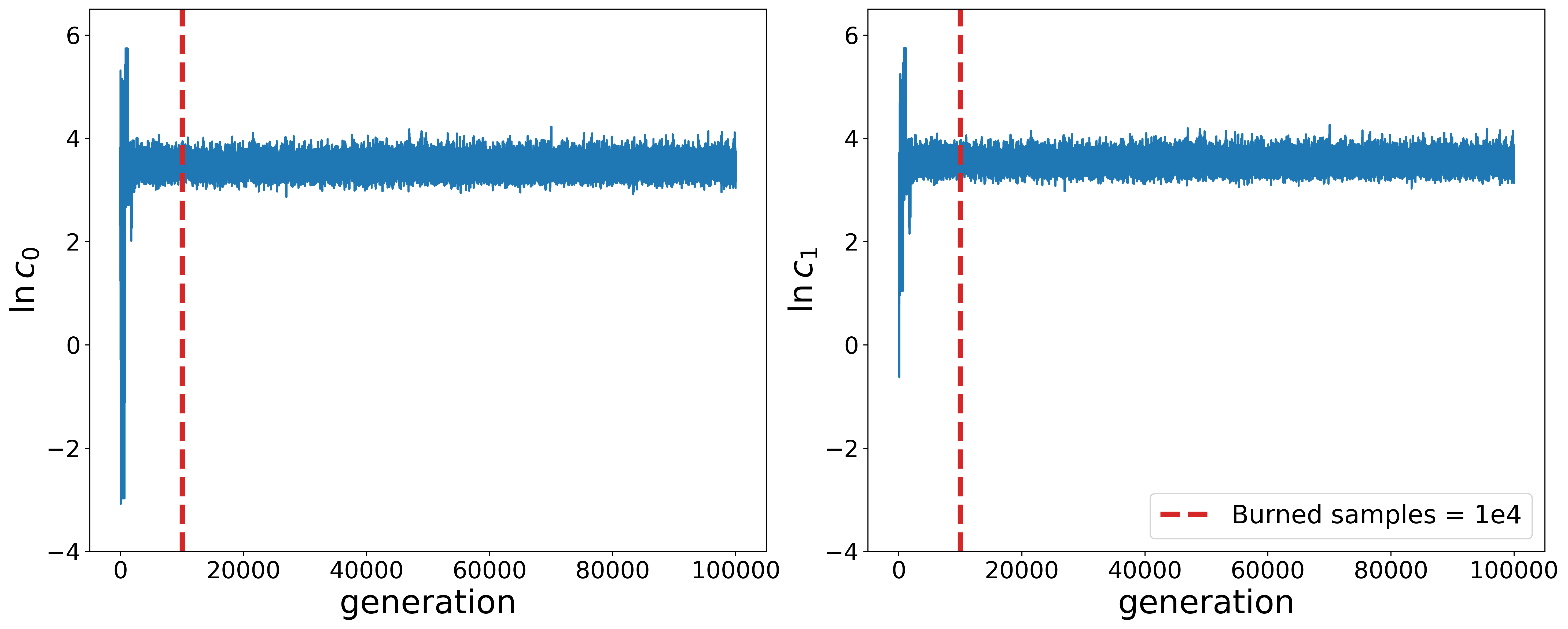}
\caption{Trace plot showing the accepted samples for $\ln c_0$ and $\ln c_1$ with the DE-MC generation.}
\label{fig:trace}
\end{figure}

\begin{figure}[ht]
\centering
\includegraphics[width=.95\textwidth]{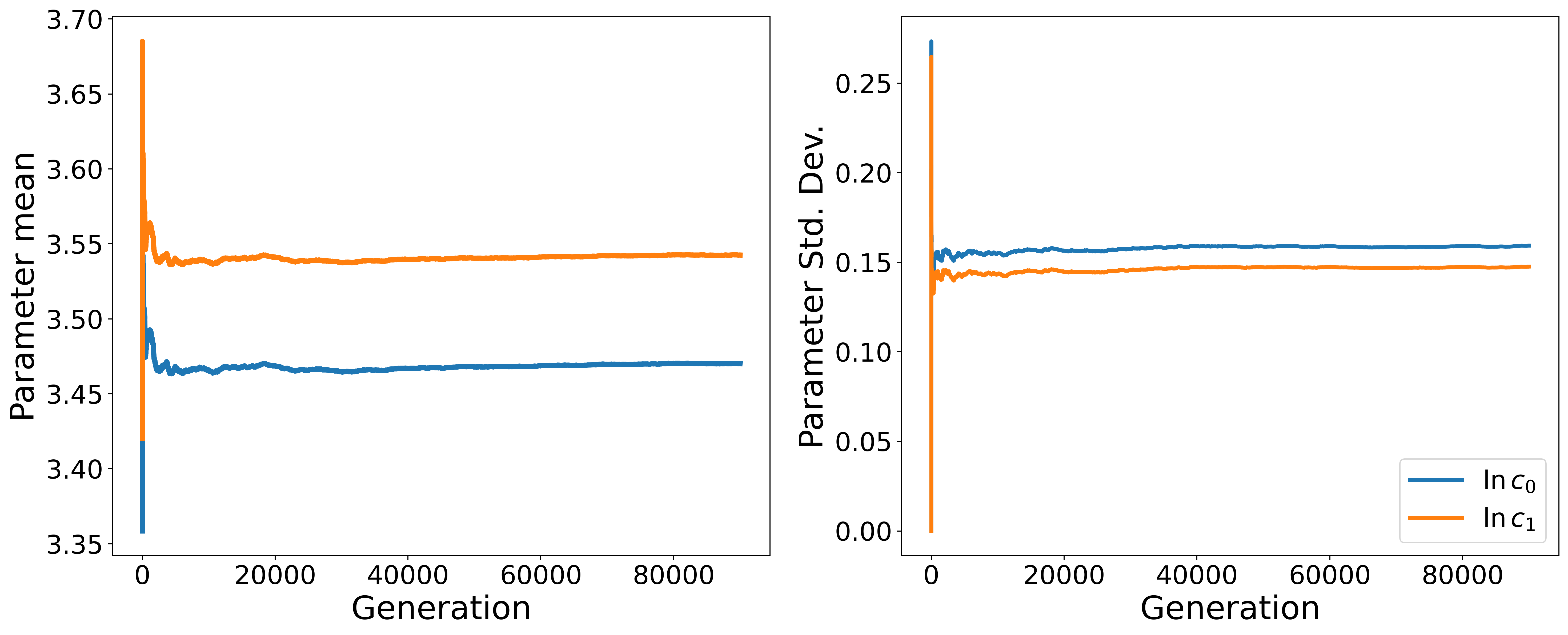}
\caption{Parameter mean and $\sigma$ convergence with number of DE-MC generation.}
\label{fig:convergence}
\end{figure}

A prior distribution was chosen as $\mathcal{N}(0.0, 2.0^2)$ for each parameter and it was found that the marginal parameter posterior distributions are contained within the prior. 
Choosing the prior distribution in this way assumes that the parameters are independent, however, it can be seen from the estimated posterior covariance matrix, $\boldsymbol{\Sigma}$, and the pair plots in Fig.\ \ref{fig:pairplot} that the DE-MC sampling captures the covariance of the model parameters, which are highly correlated. 
This parameter correlation can explain why initial attempts to sample the parameters failed to converge using traditional MCMC. 
However, the trace plots shown in Fig.\ \ref{fig:trace} and calculating the parameter means and standard deviations as a function of the number of DE-MC generations in Fig.\  \ref{fig:convergence} shows that the DE-MC sampling is well-converged for both parameters. 
Moreover, the sampling acceptance rate is $> 30\%$, which is within the expected range for a multivariate normal target with dimension $d=2$ \cite{cajo_j_f_ter_braak_markov_2006}. 

\subsection{Model fit and uncertainty quantification}

\begin{figure}[h]
\includegraphics[width=.95\textwidth]{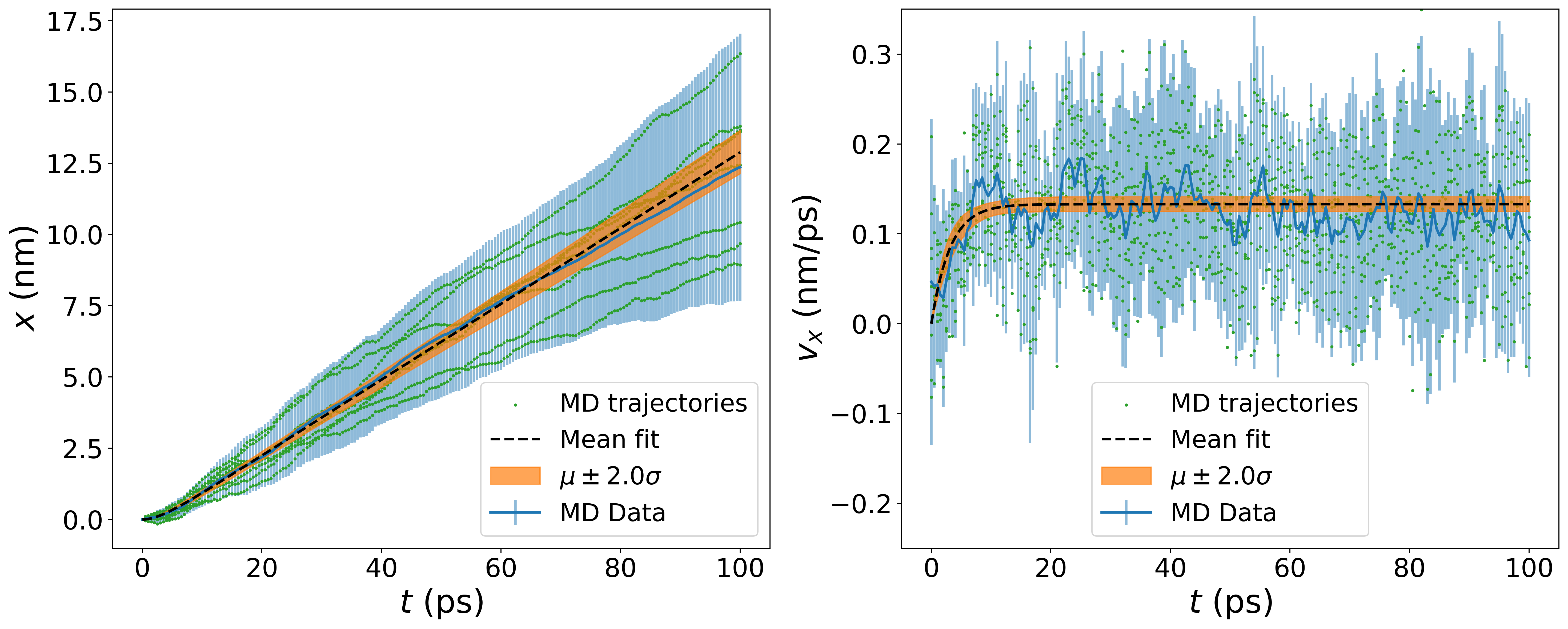}
\caption{Model fit to the MD dislocation trajectories (position: left, velocity: right). The raw MD data is shown in green, and the mean dislocation positions and velocities at time $t_i$ are shown in blue with the error bars representing $\pm$ 2 standard deviations from the mean.  
The mean fit is represented by the dashed black line, obtained by evaluating the model at the mean parameter values calculated from the DE-MC sampling. 
The orange shaded areas represent the uncertainty in the dislocation positions and velocity predicted by the model obtained by sampling the posterior distribution determined using DE-MC.}
\label{fig:dim_fit}
\end{figure}

Samples were drawn from the joint posterior distribution given in (\ref{eqn:estimated_posterior}). 
For each sample, the corresponding $B/m$ and $F/m$ were calculated using the expressions in (\ref{eqn:non_dim_params}). 
The samples were then used to evaluate the model at each timestep $t_i$, where the equation of motion in (\ref{eqn:dim_model_1}) was solved numerically to give both the position, $x$, and the velocity, $v_x$ ($dx/dt$). 
This gives a distribution for the position, $x_i$, and velocity, $v_{x_i}$, at every $t_i$. 
The mean values of $m$, $B$, and $F$ were used to produce the mean fit shown as the dashed black line in Fig.\ \ref{fig:dim_fit} with an RMSE of \SI{0.2}{\nano\metre}. 
In Fig.\ \ref{fig:dim_fit}, the shaded regions in orange represent the uncertainty in the positions and velocities predicted by the model, which was calculated as $\pm \,2 \,\sigma$, where $\sigma$ is the standard deviation of the distribution of positions and velocities at each time step $t_i$.

The terminal dislocation velocity $v_t$, was determined to be \SI{0.13}{\nano\metre\per\pico\second} from the mean fit at $t=\SI{100}{\pico\second}$. 
Using $v_t$, the applied shear stress $\tau = \SI{10}{\mega\pascal}$, and $|\mathbf{b}| = \frac{\sqrt{2}}{2} a_0$, the damping coefficient, $B$, was calculated using equation \eqref{eqn:B} to be \SI{18.72}{\pico\newton\ps\per\nm\squared}. 
The values obtained for $v_t$ and $B$ compare well to those in literature \cite{bitzek_atomistic_2004} for the conditions used in our simulations. 

Using the mean values of $c_0$ and $c_1$ obtained from the marginal parameter distributions, the corresponding values of $B/m$ and $F/m$ were calculated. 
These were then used with the calculated value for $B$ to calculate $m = \SI{58.25}{\pico\newton\ps\squared\per\nm\squared}$ and $F = \SI{2.49}{\pico\newton\per\nm}$, which corresponds to the \SI{10}{\mega\pascal} shear stress imposed on the system. A large discrepancy was found between the mean value of $m$ calculated from the fitting procedure presented here and that in Bitzek et al.\ \cite{bitzek_atomistic_2004}. However, a direct comparison of dislocation mass is not straightforward due to its dependence on factors such as the simulation box size. The theoretical edge dislocation rest mass $m_0^{\perp}$ can be calculated using the expression from Sakamoto \cite{sakamoto_high-velocity_1991}:
\begin{equation}
    m_0^{\perp} = \left(1+ \frac{c_t^4}{c_l^4}\right)\frac{\mu b^2}{4\pi c_t^2}\ln\left(\frac{R}{r_0} \right)
    \label{eqn:rest_mass} 
\end{equation}
where $c_t$ and $c_l$ are the transverse and longitudinal wave speed, respectively; $\mu$ is the shear modulus; and $R$ and $r_0$ are the upper and lower integration limits, respectively. Based on recommendations from \cite{john_price_hirth_theory_1982, d_hull_introduction_2011}, setting $R$ as the distance between the dislocation and the first fixed atomic layer in our simulation box and $r_0 = 4b$, we get a theoretical $\frac{m_0^{\perp}}{2} = \SI{58.47}{\pico\newton\ps\squared\per\nm\squared}$, where the factor of $\frac{1}{2}$ accounts for the dissociation of the dislocation into two partials \cite{bitzek_atomistic_2004}. Good agreement is found between the mean $m$ calculated from the fitted parameters and this theoretical value. This agreement and the values obtained for $v_t$ and the corresponding $B$, which recover the applied shear stress, gives us confidence in the fitting procedure and the suitability of the model used.

\begin{figure}[h]
\centering
\includegraphics[width=.55\textwidth]{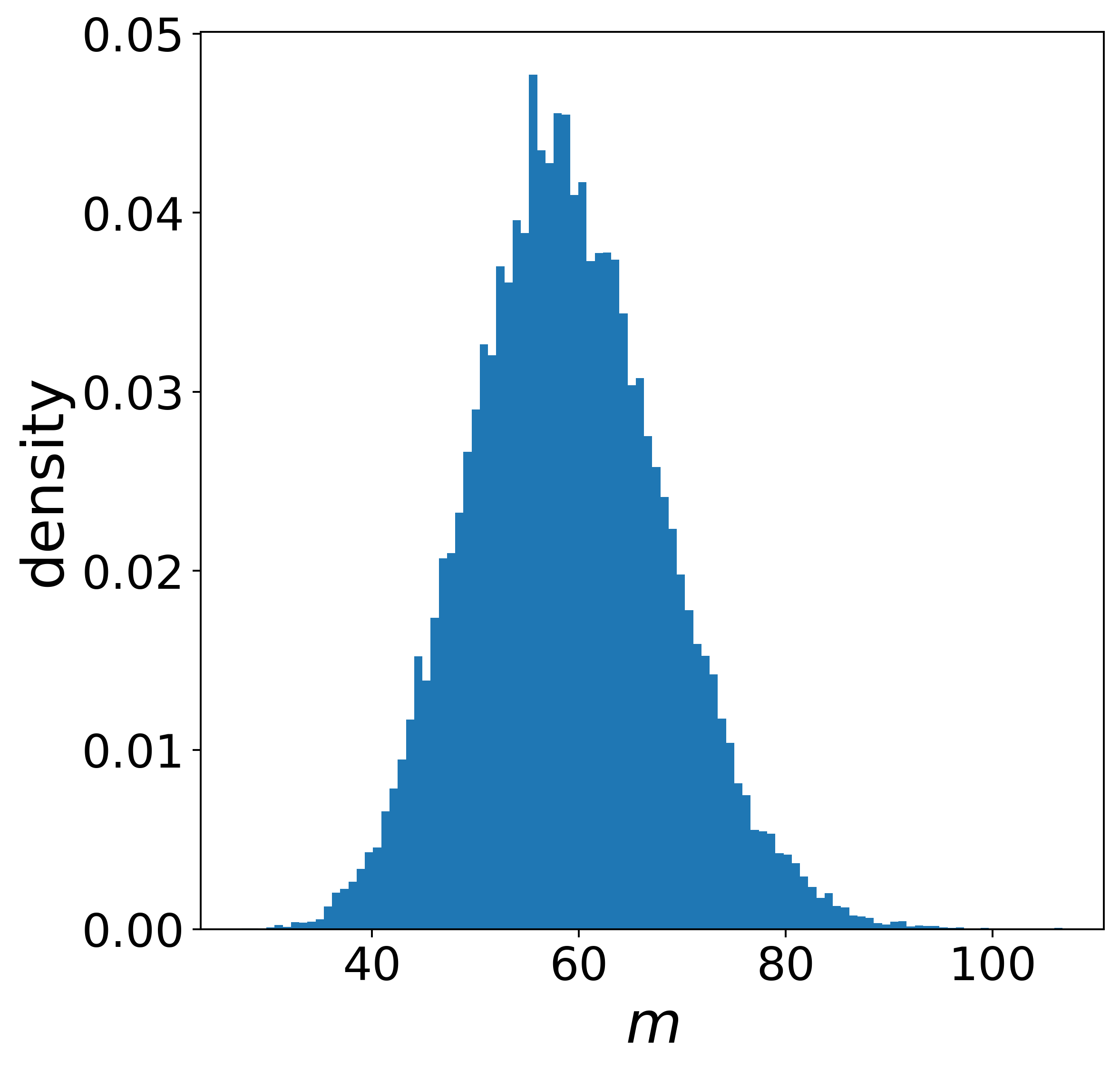}
\caption{Distribution of the dislocation effective mass, $m$, calculated by dimensionalising the accepted samples of $\ln c_0$ and assuming B = \SI{18.72} {\pico\newton\pico\second\per\nano\meter\squared} calculated from the terminal dislocation velocity determined from the mean fit}.
\label{fig:m_conditional}
\end{figure}

While the use of a sampling approach was intended to also give uncertainty bounds on the individual model parameters, it should be noted that due to the high degree of correlation between $c_0$ and $c_1$, it does not seem reasonable to give the uncertainty of the values for $m$, $B$ and $F$ as a simple $\pm 2 \sigma$.
This would lead to a significant overestimation of the uncertainty in simulation results using parameters sampled independently using those error bounds. 
The full uncertainty is encapsulated in the covariance matrix \eqref{eqn:estimated_posterior} and its effects are graphically represented in Fig.\ \ref{fig:dim_fit}, where the uncertainty in the parameters is propagated through the model to give uncertainties on the dislocation positions and velocities predicted by the model. However, if the value of B is fixed at that calculated using $v_t$ as determined from the mean fit, the accepted samples of $\ln{c_0}$ can be transformed using the expression for $c_0$ in (\ref{eqn:non_dim_params}) to get a distribution for m shown in Fig.\ \ref{fig:m_conditional}. 
This distribution shows that variance of $m$ is large, making it difficult to know with certainty.

The uncertainty in the model predictions determined by sampling the posterior distribution is shown as the orange shaded region in Fig. \ref{fig:dim_fit}. 
The uncertainty estimated in this way describes the deviation of the mean dislocation position from the mean fit well, however, it does not capture the noise in the dislocation position data determined from the different simulation runs (blue shaded region in \ref{fig:dim_fit}. 
A possible explanation for this is our assumption that the noise is uncorrelated in time, which can lead to an underestimation of the uncertainty in the model predictions \cite{carter_parameter_2009, cubillos_correlated-noise_2016}. 
This will be considered as the model is further developed in a future work.

\subsection{Current model limitations}

While the method described above was able to obtain consistent values for $m$, $B$, and $F$, the approach has some limitations. 
In the following, we present two cases that cannot be adequately described by the simple equation of motion, but which are important for modelling material plasticity, namely the case of multiple interacting dislocations, and a dislocation interacting with an obstacle in the form of a precipitate. 

\subsubsection{Dislocation interactions with long-range elastic field defects} 
So far, we have only considered modelling the trajectory of a single dislocation in pure Ni. 
Dislocations are line defects with a long-range elastic field and having multiple dislocations means that their elastic fields interact. 
Material deformation is governed by the motion of a large number of dislocations and accordingly, realistic modelling of material deformation behaviour requires considering dislocation-dislocation interactions. 

In FCC materials, $\frac{a}{2}\langle 110\rangle\{111\}$ edge dislocations are perfect dislocations, meaning they can move through the material without disrupting the lattice. 
In Ni-based superalloys, dislocations of this type move from the Ni solid solution matrix (modelled here as pure Ni) and can enter the $\gamma'$ precipitates. 
Since $\gamma'$ precipitates are Ni$_3$Al with an L1$_2$ structure, dislocations of this type create an anti-phase boundary (APB) when passing through them. 
Accordingly, $\frac{a}{2}\langle110\rangle\{111\}$ dislocations move in pairs, where the leading dislocation creates the APB and the trailing dislocation restores the crystal structure.

In addition to dislocations travelling in pairs, another important dislocation-dislocation interaction that should be accounted for in Ni-based superalloys occurs when there are misfit dislocations forming at the interface of incoherent precipitates \cite{zhu_misfit_2005}. 
Due to the difference in lattice constants between the $\gamma$ (\SI{0.352}{\nm}) and $\gamma'$ (\SI{0.357}{\nm}) phases \cite{mishin_atomistic_2004}, if the interface between these two phases is large enough, misfit dislocations form along the interface. 
Accordingly, dislocations from the $\gamma$ matrix encountering incoherent precipitates would also interact with these misfit dislocations.

\subsubsection{Dislocation-precipitate interactions}

In the pure Ni cell, the dislocation remains relatively straight throughout the simulation. 
However, in the presence of an obstacle, the dislocation can loop around it or it can bend if it travels through the obstacle, depending on the strengthening mechanism \cite{goodfellow_strengthening_2018}. 
In Ni-based superalloys, precipitates play a major role in determining the material's deformation behaviour by hindering dislocation motion and leading to the extraordinary strengthening effect they exhibit, especially at high temperature. 
In addition to the dislocation-dislocation interactions described previously, any model used to describe the deformation behaviour of Ni-based superalloys needs to also be able to capture dislocation bending due to interactions with precipitates.

\section{Conclusions}\label{sec:conclusions}
In this paper, we have presented a methodology which uses DE-MC as a sampling technique to obtain the distributions of the parameters $m$, $B$, and $F$ of an equation of motion by fitting to MD dislocation trajectories. 
Using the methods described, we are able to obtain mean values of $B$ and $v_t$ that are comparable to those in literature for similar simulation conditions. 
Moreover, our method recovers the value of the applied shear stress that was used in the simulation. 
The main advantage of fitting the parameters using sampling is being able to determine the dislocation positions and velocities as a function of time while quantifying the uncertainties in the model predictions.

Based on the above, we believe that the model and fitting procedure described here work well for predicting the dislocation dynamics of straight dislocations in the absence of interactions with other dislocations and obstacles such as precipitates. 
We have discussed how such interactions must be described for more complex systems such as Ni-based superalloys, which is our main system of interest. 
This work is a first step towards building a more predictive surrogate model that is able to capture the deformation behaviour of these alloys given inputs such as precipitate size, arrangement, and morphology.

\section*{Data availability}
The data that support the findings of this study are openly available at  \cite{geraldine_anis_dataset_nodate}. Code used for analysis is also openly available at \url{https://github.com/geraldineanis/DislocCode.git}.

\section*{Acknowledgements}
This work was funded and supported by the EPSRC Centre for Doctoral Training in the Modelling of Heterogeneous Systems (EP/S022848/1), with PhD scholarship (GA) being gratefully acknowledged. The work was performed using computational resources from the Scientific Computing Research Technology Platform (SCRTP) at the University of Warwick and the Sulis Tier 2 HPC platform hosted by SCRTP. Sulis is funded by EPSRC Grant EP/T022108/1 and the HPC Midlands+ consortium. GA acknowledges mentoring and support from Prof.\ Julie Staunton, University of Warwick.

 \bibliographystyle{elsarticle-num} 
 \bibliography{references.bib}

\end{document}